# Constrained motion of self-propelling eccentric disks linked by a spring

Tian-liang Xu[1], Chao-ran Qin[1], Bin Tang[1], Jin-cheng Gao[1], Jiankang Zhou[2*], Kang Chen[1*], Tian Hui Zhang[1*], Wen-de Tian[1*]

[1] *Center for Soft Condensed Matter Physics and Interdisciplinary Research, Soochow University, Suzhou 215006, China*

[2] *School of Optoelectronic Science and Engineering, Soochow University, Suzhou 215006, China*

Email: tianwende@suda.edu.cn kangchen@suda.edu.cn, zhangtianhui@suda.edu.cn, health@suda.edu.cn.

**Abstract:**

It has been supposed that the interplay of elasticity and activity plays a key role in triggering the non-equilibrium behaviors in biological systems. However, the experimental model system is missing to investigate the spatiotemporally dynamical phenomena. Here, a model system of an active chain, where active eccentric-disks are linked by a spring, is designed to study the interplay of activity, elasticity, and friction. Individual active chain exhibits longitudinal and transverse motion, however, it starts to self-rotate when pinning one end, and self-beats when clamping one end. Additionally, our eccentric-disk model can qualitatively reproduce such behaviors and explain the unusual self-rotation of the first disk around its geometric center. Further, the structure and dynamics of long chains were studied via simulations without steric interactions. It was found that hairpin conformation emerges in free motion, while in the constrained motions, the rotational and beating frequencies scale with the flexure number (the ratio of self-propelling force to bending rigidity), $\chi$, as $\sim(\chi)^{4/3}$. Scaling analysis suggests that it results from the balance between activity and energy dissipation. Our findings show that topological constraints play a vital role in non-equilibrium synergy behavior.

## 1.Introduction

Active systems, composed of active agents which convert internal energy or energy from environment for directed motion, exhibit exceptional non-equilibrium properties resulting in collective motion, anomalous dynamical behaviors. [1–11]. Polymer-like active agents are rich in nature at various scales such as microtubules driven by motor proteins[12,13], slender bacteria[14] and macroscopic worms[15,16]. Microtubules can form a ring structure and self-organize into a lattice of vortices[17,18]. Similarly, microfilaments can form vortices, clusters and banded structures[12]. In addition, active colloidal chains were employed to mimic living matters and explore novel active structures [19–21]. These kinds of active chains can produce periodic oscillations and synchronization behaviors at the single chain level[21].

Inspired by biological systems, a model of self-propelling filament (SPF) were proposed [22]. The SPF is a chain with active force acting tangentially to the contour of chain regardless of its shape [23–28], which is a kind of follower-forces. The self-beating and flapping of SPF arises from the 'follower force'-induced buckling instability[29–31] due to its intrinsic non-variation[32–34]. Experimentally, Zheng et al.[35] design an elastoactive structure via embedding Hexbug in the laser-cut silicon rubber chain at a centimeter-sized scale. They found such structures exhibit flagellar motion when pinned at one end, self-snapping when pinned at two ends, and synchronization when coupled together with a sufficiently stiff link. Despite these advances, there are as of yet few model experimental platforms in macroscopic scale to study the constrained motion of active agents. It is interesting to explore the effect of topological constraint on the collective behavior of active agents.

Here, we experimentally design an active chain via connecting five self-propelling eccentric disks by a spring and combine computer simulations to investigate its structure and dynamics. Our previous work[36] found that the eccentricity leads to collective motion of disks via collision due to the mutual coupling of the positional and orientational degrees of freedom. We find that the polar state of disks also occurs after connected by spring. For free motion, the chain shows two modes of motion (i.e., longitudinal and transverse motion) on the dependence of translational and rotational friction coefficient. Furthermore, the chain rotates when one end was pinned and periodically beats when the end was clamped. Our simulations of eccentric-disk model can qualitatively reproduce such behaviors and explain the unusual self-rotation of the first disk around its geometric center. We also extended to the long chain systems via computer simulations and found the angular and beating frequencies scale with the flexure number, $\chi$, as $\sim(\chi)^{4/3}$. Overall, our experiment and model provide a novel view for the interplay of activity, elasticity, and friction.

This paper is organized as follows. The second section presents the experiment setup and the computational model as well as the method. In the third section, the spatiotemporal dynamics and its mechanism of self-propelling eccentric-disk chain are systematically analyzed under various conditions. Finally, a summary is given.

## 2. Experimental design and simulation methods

**Experimental setup:**

An active chain is made by connecting five self-propelling disks with a spring as illustrated in Fig.1a. The disk is made of microrobot (Hexbug Nano) and a laser-cut circular foam plate (Fig.1a). The self-propelling behavior of Hexbug Nano is realized by its soft rubber legs interacting repetitively and impulsively with the ground, which can convert the vibrations into directional motion. The disk is eccentric because the tail of microrobot is heavy. There are two white spots with a diameter of 13mm and 9mm, respectively, on the disk, to track its position and orientation. The behavior of the self-propelling disk is given in Supporting Information (SI). The geometric centers (GCs) of five disks with the interval of 6.4cm were pinned on a short spring by pushpins, which could ensure the disks rotating freely (Fig.1a). Springs with rigidities, about 7.01x10$^{-4}$ $Nm^2/rad$ (weak bending stiffness) and 3.51x10$^{-3}$ $Nm^2/rad$ (strong bending stiffness), were used to study the interplay of elastic effect in experiments. To systemically investigate the cooperative behavior of self-propelling disks, three boundary conditions were designed: free movement (free mode), one terminal of spring pinned (pinning mode), and one terminal clamped (clamping mode).

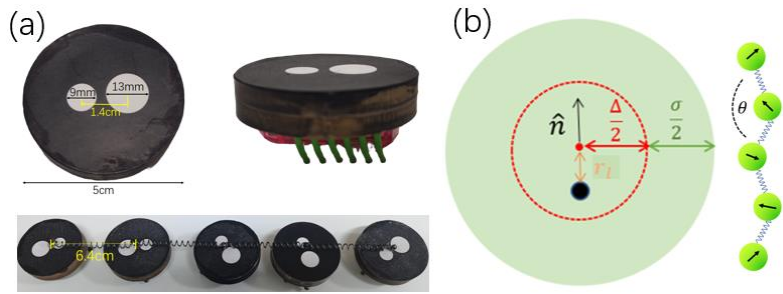

Fig.1 (a) The top and side view of the self-propelling eccentric disk composed of a Hexbug Nano and a circular foam plate, and the active chain with 5 self-propelling disks connected by an elastic spring. The white spots with diameter of 9mm and 13mm was used to track the position and orientation of each disk with diameter of 5cm. The distance of adjacent disks is 6.4cm. (b) Schematic diagram of disk model and the active chain for simulation. The red point denotes the geometric center (GC) of the disk, while the black point denotes its center of mass (COM). The disk is self-driven along the orientation, $\hat{\boldsymbol{n}}$, from COM to GC. The dotted circle denotes a hard core with diameter, $\Delta$. $\theta$ is the bending angle, $r_l$ the distance between the COM and GC.

**Simulation model and methods:**

To mimic the self-propelling eccentric disk, a circular rigid body with two centers was adopted: one is the geometric center (GC), and the other is the center of mass (COM). The distance between the COM and GC is $r_l = 1/3\sigma$. A disk $i$ is described by its geometric center (GC), $\boldsymbol{r_i(t)}$, center of mass (COM), $\boldsymbol{r_{ci}(t)}$, and orientation, $\boldsymbol{\hat{n}_i(t)}([\cos\phi, \sin\phi])$, which goes from COM to GC, as demonstrated in Fig.1b. All disks interact with a repulsive Lennard-Jones potential, with the distance shifted by $\Delta$.

$$U(r_{ij}) = \begin{cases} 4\varepsilon\left[\left(\dfrac{\sigma}{r_{ij}-\Delta}\right)^{12} - \left(\dfrac{\sigma}{r_{ij}-\Delta}\right)^{6}\right] & r_{ij} < \Delta + \sqrt[6]{2}\sigma \\ 0 & otherwise \end{cases} \quad (1)$$

where $r_{ij}$ is the distance of GCs between disk $i$ and $j$.

The harmonic potential was adopted to connect the adjacent disks, $U_b = K(r - r_0)^2$, where the bond constant $K$ is 10000 $\frac{\varepsilon}{\sigma^2}$. $r$ refers to distance of GCs, $r_0 = 2.55\sigma$, the equilibrium length. The bending of spring is modeled by the potential, $U_a = \kappa_\theta(\theta - \pi)^2$, with bending angle, $\theta$, and rigidity coefficient, $\kappa_\theta$, which is a parameter in our simulations (Fig.1b).

The dynamics of each disk is described by the following equations:

$$m_c \ddot{\boldsymbol{r}}_{ci}(t) = F_a \hat{\boldsymbol{n}}_i(t) - \nabla U(r) - \gamma_t \dot{\boldsymbol{r}}_{ci}(t) + \sqrt{4\gamma_t k_B T_t} \boldsymbol{\eta}_i(t)$$

$$I \ddot{\phi}_i(t) = \Gamma_i - \gamma_r \dot{\phi}_i(t) + \sqrt{2\gamma_r k_B T_r} \boldsymbol{\xi}_i(t) \quad (2)$$

$$\Gamma_i = r_l \; \hat{\boldsymbol{n}}_i \otimes [-\nabla U(r)]$$

Where $m_c$ is the mass of COM, $I$ the rotary inertia, $\gamma_t$ the translational friction coefficient, $\gamma_r$ the rotational friction coefficient, $F_a$ the strength of self-propelling force, $k_B$ the Boltzmann constant. $T_t$ and $T_r$ are strength of translational and rotational noises, respectively. $U(r)$ includes pair, bond, and angle interactions of the whole system. For the granular system, the $T_t = T_r$ is not mandatory. $\Gamma_i$ denotes the torque on disk $i$. Besides, $\boldsymbol{\eta}_i(t)$ and $\boldsymbol{\xi}_i(t)$ are Gaussian white noises on the COM with zero mean and unit variances, as follows.

$$\langle \eta(t) \rangle = 0; \langle \eta_\alpha(t) \eta_\beta(t') \rangle = \delta_{\alpha\beta} \delta(t - t')$$

$$\langle \xi(t) \rangle = 0; \langle \xi(t) \xi(t') \rangle = \delta(t - t') \quad (3)$$

where $\eta_\alpha$ or $\eta_\beta$ is the x/y-th component of the vector $\eta(t)$.

We used the home-modified LAMMPS software to perform all simulations. The equations of motion were solved using a velocity-Verlet algorithm. Periodic boundary conditions were applied in the X and Y directions in two dimensions (2D). Reduced units were used by setting $m=1$, $\sigma=1$, and $\varepsilon=1$. The corresponding time unit, $\tau = \sqrt{m\sigma^2/\varepsilon}$. Additionally, we set $m_c=1.1m$, $I=0.01m\sigma^2$, the reduced translational noise, $k_B T_t = 1.0\varepsilon$, shift distance, $\Delta = \sigma$, $L=100$ $\sigma$, and the unit of $\gamma_t$, $\sqrt{m\varepsilon/\sigma^2}$. For each case, it ran in a minimum time of $1 \times 10^4 \tau$ with a time step, $10^{-4}\tau$. The chain length, $N$, is 5 in our simulation for comparing with our experiment, if not special statement.

## 3. Result and Discussion

### 3.1 Free chain

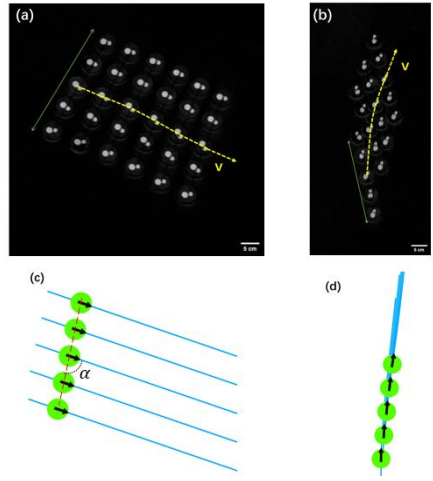

Fig.2 Time-lapse snapshots showing transverse (a) and longitudinal (b) motions of the active chain of strong stiffness in experiments. The dotted yellow arrow indicates the direction of the movement, the double green arrow represents the end-to-end vector of spring. The corresponding snapshots of transverse (c) and longitudinal (d) motions obtained by simulations with parameters $\gamma_r = 100$ and $\gamma_r = 1$ respectively, at $F_a = 500, \kappa_\theta = 100, T_t = T_r = 0$. The blue line is the trajectory of the disks and the black arrow is the orientation of each disk. $\alpha$ is the average angle between the direction of centroid velocity and the end-to-end vector of the chain.

We first investigate the behavior of active chains freely moving on the black tempered glass. Two modes of motion were observed for the semi-flexible chain: transverse and longitudinal motions (Fig.2). The chain moves transversely with the direction of centroid velocity nearly perpendicular to its contour (Fig.2a, Move_1 in SI) or moves longitudinally with the direction of centroid velocity almost along its contour (Fig.2b, Move_2 in SI). One mode could switch to the other possibly due to the roughness of ground and the noise of motor vibration. The behavior is similar for the case of weakly-stiff chain (data not shown), which implies that the behavior slightly depends on the stiffness. To confirm the existence of two modes, we perform numerical simulations without noise via setting $T_t = T_r = 0$. Typical trajectories are given in Fig.2c and Fig.2d .The similar behaviors are, indeed, observed.

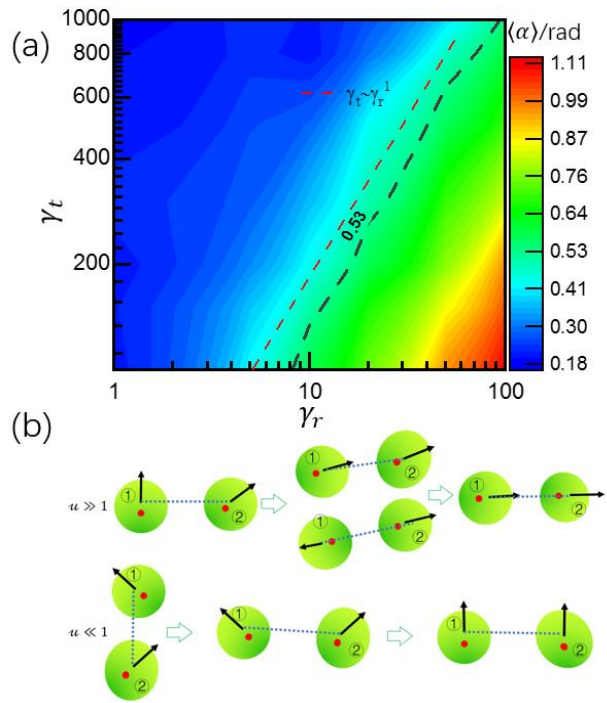

Fig.3 (a) $\gamma_t$-$\gamma_r$ phase diagram of two motion modes. The color bar from blue to red denotes the value of $\langle\alpha\rangle$ from 0 to 1.1. Here it should be noted that we set $\alpha = \pi - \alpha$ when $\alpha > \pi/2$. The dotted line denotes the scaling law of $\gamma_t \sim \gamma_r^1$. (b) Schematic diagram of rotation and translation-induced alignment of disks.

To understand the behavior and explore what determines the dominant motion mode, we study the dependence of $\langle\alpha\rangle$ on $\gamma_t$ and $\gamma_r$ at $T_t = T_r = 0$, by simulations. $\langle\alpha\rangle$, which was calculated by 1000 independent samples, denotes the average angle between the direction of centroid velocity and the end-to-end vector of chain. It should be noted that we set $\alpha = \pi - \alpha$ when $\alpha > \pi/2$ in order to calculate $\langle\alpha\rangle$. $\langle\alpha\rangle \sim 0$ means the longitudinal motion, $\langle\alpha\rangle \sim \pi/2$ transverse motion. The centroid velocity is the average velocity of five disks. $\gamma_t$ ($\gamma_r$) determines the characteristic time of how fast the position (orentation) changes. The $\gamma_t$-$\gamma_r$ phase diagram is plotted in Fig.3a. It can be found that the large $\gamma_t$ and small $\gamma_r$ leads to small $\langle\alpha\rangle$s while the small $\gamma_t$ and large $\gamma_r$ gives rise to large $\langle\alpha\rangle$s. That is to say, the motion mode is mainly determined by the dimensionless parameter $u = \gamma_t \sigma^2/\gamma_r$. Intuitively, when $u \gg 1$, the position of each disk's COM is hardly changed, the spring's stress is mainly released via changing disks' orentation, which prefers to the longitudinal motion. Conversely when $u \ll 1$, the orentation of each disk is hardly altered, the stress release is mainly through translation of COM, then the average orentaiton along the contour is low probabilistic, ultimately the transverse motion takes place. The physical intuition can be uncovered by a dimer model (Details are given in SI).

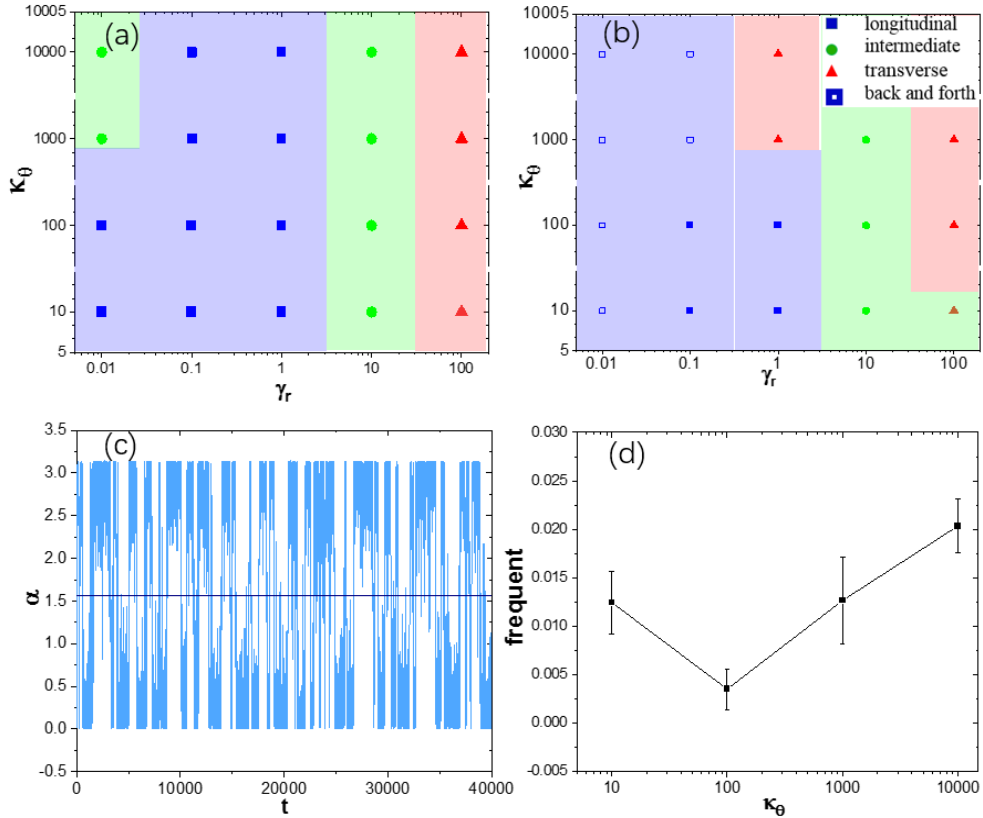

Fig.4 Phase diagram of motion mode at (a) $T_t = T_r = 0$ and (b) $T_t = 1.0, T_r = 0.0$. (c)Time evolution of $\langle\alpha\rangle$ and (d) reverse frequency of longitudinal motion as a function of chain rigidity for system at $\gamma_r = 0.01$, $\kappa_\theta = 1000$, $T_t = 1.0, T_r = 0.0$. Here all simulations are performed at $F_a = 500$.

We further explore the effect of translational noise and stiffness on chain motion by simulations. To quantitatively distinguish the motion modes, $\langle\alpha\rangle < \pi/6$ is considered to be longitudinal mode, $\langle\alpha\rangle > \pi/3$ transverse mode, and other intermediate state. Phase diagrams without/with translational noise are shown in Fig.4a and Fig.4b, respectively. It can be observed that, at the same $\gamma_r$, the increase of rigidity does not cause the change of mode significantly in the systems without noise. The system with noise is similar with the increase of rigidity. However, the distinct difference is that frequent reversal of longitudinal motion occurs at small $\gamma_r$ due to noise-enhanced rotation of disks. The frequent reversal, which can be manifested by the time evolution of $\alpha$ (Fig4.c), leads to the back-and-forth motion of chain. This means that the noise destroys the stability of polar motion. The unique property roots in the weak correlation of disk's orientation and bonds of chain, which is not found in the SPF system. In addition, we find the frequency of back-and-forth depends on the chain stiffness as shown in Fig.4d. The frequency slightly decreases, then increases when increasing rigidity of chain. In addition, when $\kappa_\theta \geq 1000, \gamma_r = 1.0$, the noise changes the chain's motion mode from longitudinal to transverse (Fig.4b). This can be understood in terms of bond and bending energy. Bond energy exerts stress along the contour of chain, while bending energy exerts stress almost perpendicular to the contour. When $\kappa_\theta \geq 1000$, the bending energy is stronger than the bond energy (see Fig. S4), hence the force perpendicular to the contour is dominant, which causes the transverse mode.

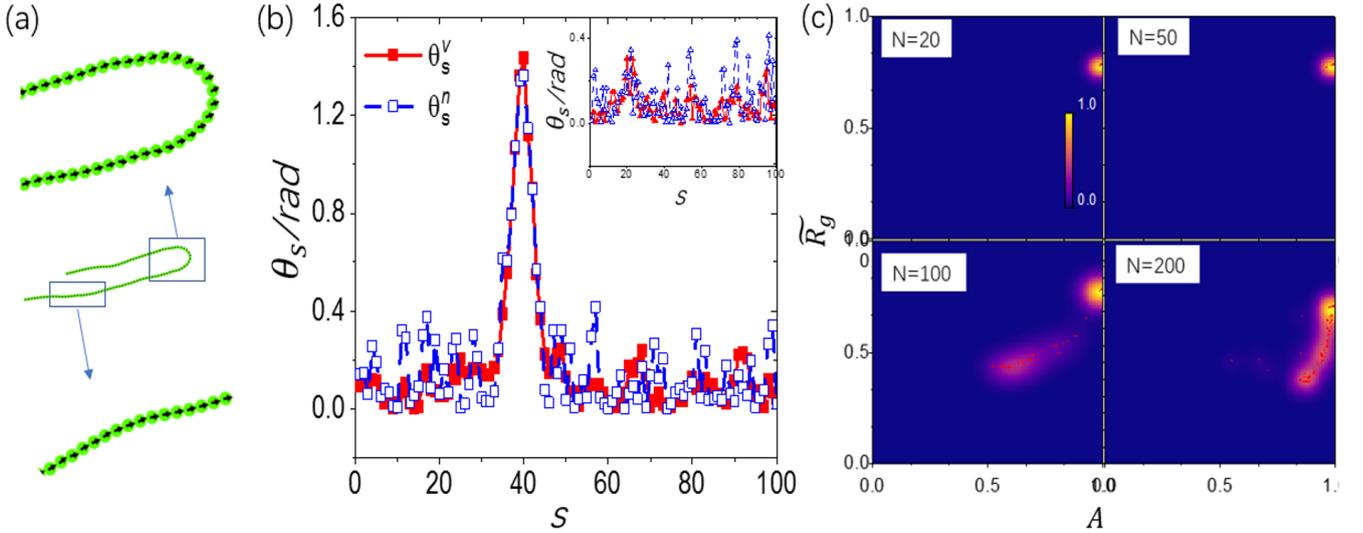

Fig.5 (a) Snapshot of the hairpin conformation with the black arrow indicating the driven direction of each disk at $F_a = 500, \gamma_t = 100, \gamma_r = 1, \kappa_\theta = 100, T_t = 1, T_r = 0$. The inset is the linear conformation. (b) The angle between velocity ($\theta_s^v$)/orientation ($\theta_s^n$) of each disk and chain contour as a function of arc-length. (c) Joint $\widetilde{R_g}$-$A$ probability density of chain conformation for various chain lengths.

Now, we focus on the chain length effect on the structure of chain in the free state at $F_a = 500, \gamma_t = 100, \gamma_r = 1, \kappa_\theta = 100, T_t = 1, T_r = 0$. The interesting hairpin conformation emerges in the long chain systems (Fig.5a). The orientations of disks in the hairpin structure are almost same, the whole chain moves with a constant speed. The angle of velocity ($\theta_s^v$) and orientation ($\theta_s^n$) of each disk relative to local contour of chain are given in Fig.5b. One can find that these angles are close to $\frac{\pi}{2}$ in the hairpin regime, 0 in the linear regime (Fig.5b). It should be noted that, at the same parameter, the polar conformation possibly emerges depending on the initial state of chain. To quantitatively characterize the appearance of hairpin and polar conformations, we extract the relative radius of gyration, $\widetilde{R_g}$, and cylindricity, $A$, from 1000 independent runs for each system (the method for calculating $\widetilde{R_g}$ and $A$ is given in SI). When they are both close to 1, the chain is an extended conformation. If they are smaller than 1, the chain is a hairpin conformation. The joint probability distributions are plotted in Fig.5c for various chain lengths. It can be found that short chains mainly display the elongated linear conformation. The hairpin conformation presents in the long chain systems with a certain probability.

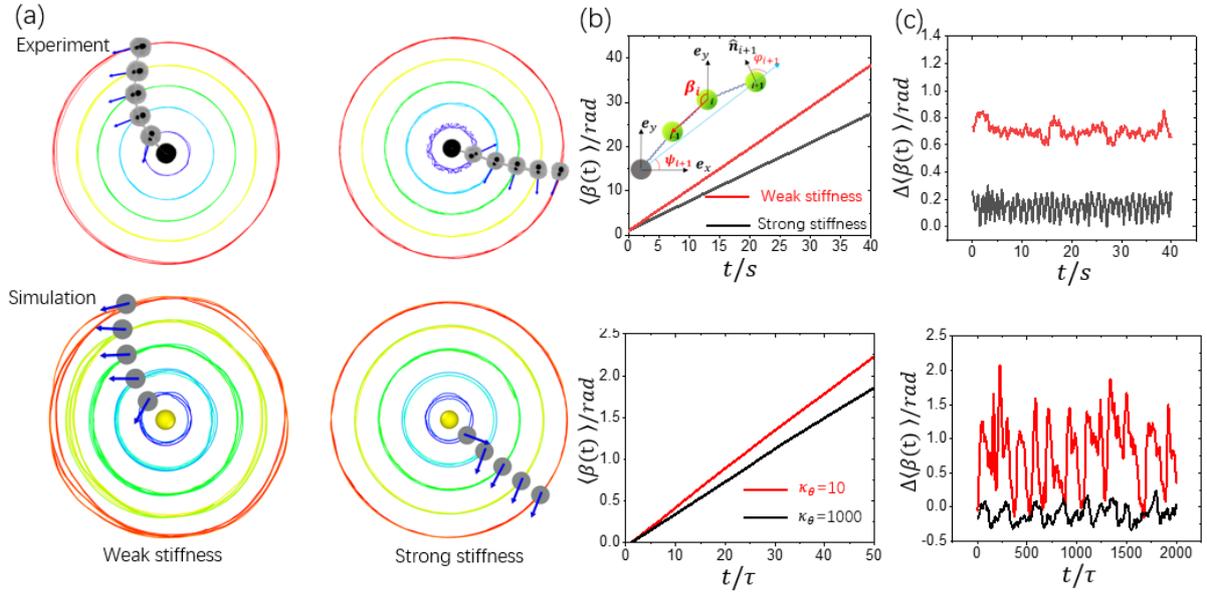

Fig.6 (a) Trajectories of disks on pinned chain with weak (left) and strong (right) rigidity, and its corresponding simulations at $\gamma_r = \gamma_t = 100$, $F_a = 100$, $T_t = 0$, $T_r = 1.0$. Circles of different colors represent the trajectories of different disks. The blue arrow denotes the self-propelling direction of each disk. (b) Time series of the mean cumulative angle $\langle \beta(t) \rangle$ for the experiments (upper) and simulations (lower). The inset is a schematic diagram. $\beta_i(t)$ is instantaneous angle between $i$th bond and $\hat{e}_y$, $\psi_i$ the angle between the vector from fixed point to the $i$th disk's geometric center and $\hat{e}_x$, $\varphi_i$ the angle between the vector from fixed point to the ith disk's geometric center and the disk's orientation. (c) Time evolution of the curvature $\Delta\beta = \beta_5 - \beta_1$ of the chain for the experiments and simulations.

## 3.2 Pinned chain

When one end of the chain was anchored on a fixed circular foam with the same height of disks, it rotates steadily with time evolution (Fig.6a, Movie_3 in SI). The final curvature of rotational chain depends on its stiffness. The simulation result is consistent with experiment. The rotational dynamics can be quantified by the mean cumulative angle $\langle \beta(t) \rangle$, where $\beta_i(t)$ is the instantaneous orientation of bond $i$ (the vector from GC of disk $i$ to the $(i-1)$th disk) with respect to vertical axis $\hat{e}_y$ (see the schematic of Fig.6b). $\beta$ linearly increases with time, indicating the chain rotates with a uniform angular speed. Strong stiffness will lower the angular speed perhaps due to the increase of rotational inertia. The mean curvature of chain is characterized by $\Delta\beta = \beta_5 - \beta_1$ (Fig.6c). More flexible chain displays larger curvature. Although our results qualitatively agree with that of pinned semiflexible SPF, the mechanism is different. The emergence of rotation for SPF is completely caused by the competition between activity and elasticity. Here the re-orientation of free rotational disks also plays a vital role in chain rotation. Imaginably, the strong stiff SPF can not rotate because all driven forces point to the fixed center without tangential component that induces a moment of force, where the emergence of rotation needs the buckling instability. That is different from our model. Our disks can freely rotate without constraints. The noise-induced orientation fluctuation will be enhanced when they are linked, then symmetry breaking and synchronous rotation appear in the constrained case. That is to say, the rotation originates from the collective motion of disks via self-arranging their propelling direction.

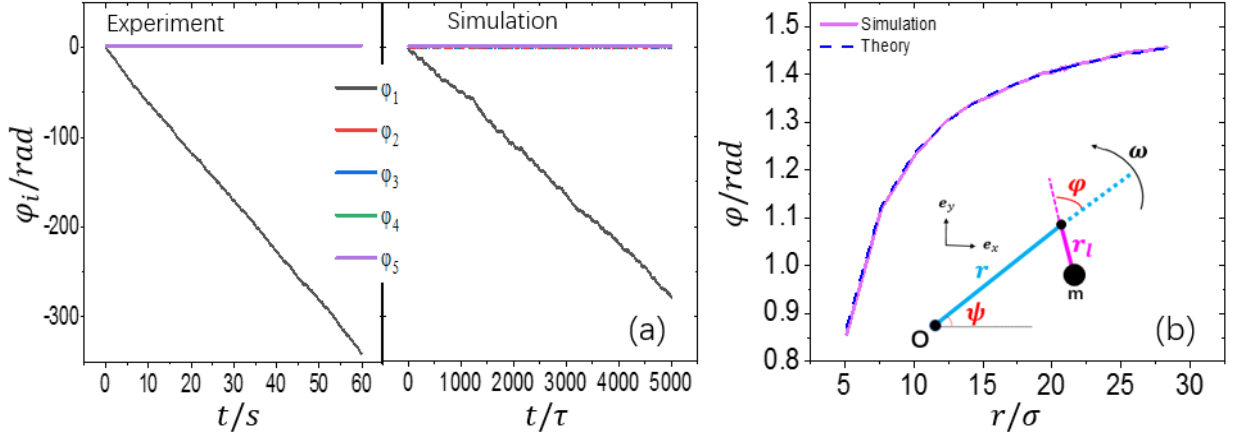

Fig.7 (a) Time evolution of $\varphi$ of each disk for experiment and simulation (b) $\varphi$ as a function of the distance, $r$, related to the fixed point. The inset is a schetch of our simplified model. The simulation is performed without steric interactions for $N = 100, \gamma_r = \gamma_t = 100, F_a = 500, \kappa_\theta = 10000, T_t = 0, T_r = 1.0$.

A surprising phenomenon emerges in our experiment and simulation: the first disk not only rotates around the fixed point, but also rotates around its GC. This can be manifested by time evolution of $\varphi_i$, the angle between the $i$th bond and the disk's orientation (see the inset of Fig. 6b). As shown in Fig.7a, $\varphi_1$ deceases with time. Besides, other $\varphi$s almost don't change with time. To understand why this happens, we construct a simplified model (A scheth is given in the inset of Fig.7b). $\psi$ is the angle between the vector from fixed point to the $i$th disk's GC and $\hat{e}_x$. $\varphi$ is the angle between the vector from fixed point to the $i$th disk's GC center and the disk's orientation. We assume the rod rotates with a fixed angular speed $\dot{\psi} = \omega$. The balance of force and moment of force gives the equation below (The detail is given in SI):

$$\dot{\varphi} = \omega \left[ \frac{r}{\frac{\gamma_r}{r_l \gamma_t} - r_l} \cos\varphi - 1 \right] \quad (6)$$

When $r < \frac{\gamma_r}{\gamma_t r_l} - r_l$, $\dot{\varphi} < 0$ always holds, implying that the self-rotational direction of disk is opposite to ω. This is why we observed the amazing behavior of the first disk in experiment and simulations. In addition, when $r > \frac{\gamma_r}{\gamma_t r_l} - r_l$, a stationary solution, $\varphi = \mathrm{acos}(\frac{\frac{\gamma_r}{r_l \gamma_t} - r_l}{r})$, is obtained via setting $\dot{\varphi} = 0$. It means that the orientation of disks only depends on the distance from the pinned point and it is perpendicular to the rod when $r \to \infty$. This is consistent with our simulation data (Fig.7b), which presents the $\varphi$ as a function of $r$ for the long chain system, $N=100$, without steric interaction.

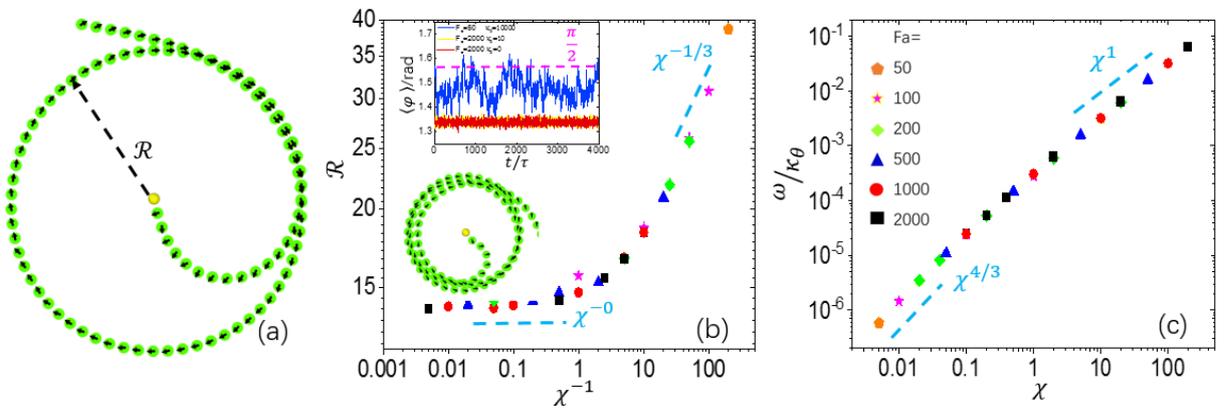

Fig.8 (a) Typical configuration of pinned chain at $N=100$, $\gamma_r = \gamma_t = 100$, $\kappa_\theta = 10000$, $F_a = 50$. (b) The radius $\mathcal{R}$ as a function of $\frac{1}{\chi}$, $\chi = F_a/\kappa_\theta$. The inset snapshot is for $\kappa_\theta = 10$, $F_a = 2000$. The inset figure shows the time evolution for various $F_a$s, and $\kappa_\theta$s (c) The rotational speed $\omega$ as a function of $\chi$.

To further understand how the interplay between activity and elasticity affects the configuration and rotation, we remove the steric interaction of disks for long chain ($N=100$) by simulations. One typical configuration is given in Fig.8a. The chain rolls into a self-overlapping circle with almost constant radius, $\mathcal{R}$, over which compression is accommodated. The size of $\mathcal{R}$ depends on the active force and rigidity of chain. We define the flexure number, $\chi = F_a\sigma/\kappa_\theta$, ratio of activity to bending rigidity. There is a master curve for the relation of $\mathcal{R}$ and $\frac{1}{\chi}$ (Fig.8b). One can find that when $\frac{1}{\chi}<1$, $\mathcal{R}$ almost does not change with $\frac{1}{\chi}$. The $\mathcal{R}$ mainly originates from the activity-induced stress along the chain. The self-driven force tends to orientate normally to the contour (see the snapshot in Fig.8b), which could keep the appearance of the circular state. This implies that, for the pinned model, even a flexible chain, it has an effective rigidity due to free-rotating active force deviating from tangential direction, which can be manifested by $\langle\varphi\rangle \sim 1.34$ (inset figure in Fig8.b). When $\frac{1}{\chi}>50$, $\mathcal{R}\sim(\frac{1}{\chi})^{\frac{1}{3}}$, which is similar to the finding of SPF system[31] due to the balancing the active force and tangential bending force in the chain. Furthermore, the rotational speed is given in fig.7c. For small $\chi$, $\omega/\kappa_\theta \sim (\chi)^{4/3}$ and large $\chi$, $\omega/\kappa_\theta \sim (\chi)^1$. This can be understood below. The linear velocity of disks v~Fa, $\omega \sim v/\mathcal{R}$, then if using the relation in Fig.8b, one can get the above scaling relation for small and large $\chi$s, respectively.

## 3.3 Clamped chain

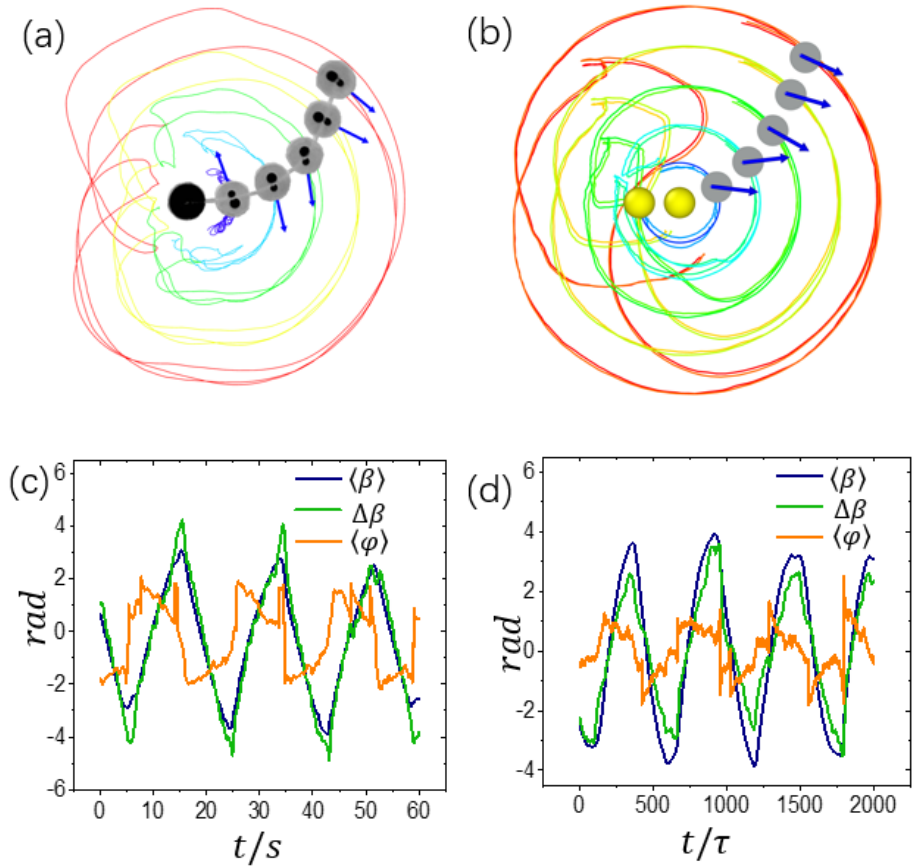

Fig.9 (a) Snapshots of trajectories of disks on the clamped chain and (b) the corresponding simulation results at $\kappa_\theta = 300$ $(left)$ with $\gamma_r = \gamma_t = 100, F_a = 100, T_t = 0, T_r = 1.0$. (c) The average swing angle $\langle\beta(t)\rangle$, the bending degree $\Delta\beta$, and the average disk orientation $\langle\varphi\rangle$ as a function of time obtained by experiment, and (d) the corresponding simulation results shows good agreement with experimental results.

When the chains were clamped, i.e., the terminal cannot freely rotate, the beating behavior emerges (Fig.9a, Movie_4 in SI) under the interplay of activity and elasticity. Our simulation results also display the similar behavior (Fig.9b). To characterize this periodic beating in experiments and simulations, we calculate the beating angle $\langle\beta(t)\rangle$, the bending degree $\Delta\beta$, and the orientation, $\langle\varphi\rangle$ relative to the chain contour, which are given in Fig.9c-d. High rigidity leads to a small beating amplitude and bending degree (See Fig.S5). The change of bending degree is almost synchronous with the beating angle, both reach peak at the same time. When the chain approaches the largest amplitude, the disk turns around quickly, thus a sharp change of $\langle\varphi\rangle$ is obtained. This non-linear behaivor is governed by activity, elasticity, and frictional damping, similar to the follower-force induced oscillations.

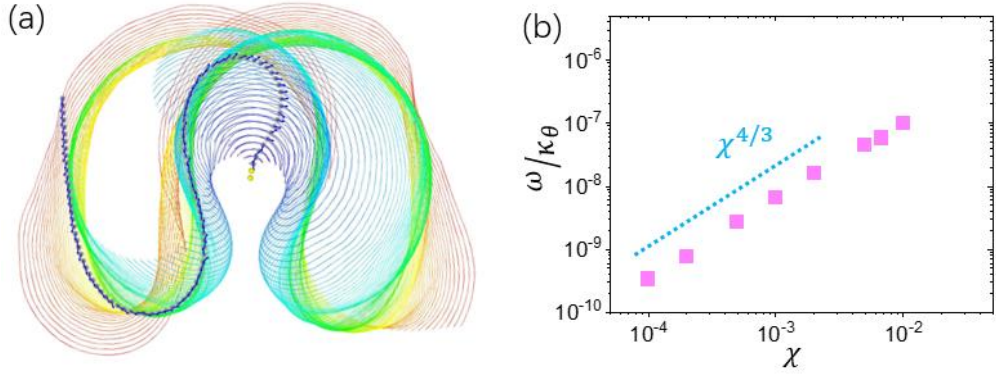

Fig.10 (a) The typical snapshot for system at same $N=100$, $\gamma_r = \gamma_t = 100, F_a = 100, T_t = 0, T_r = 1.0, \kappa_\theta = 20000$. (b) the beating frequency $\omega$ as a function of $\chi$ for $N=100$, $F_a=100$.

We further increase the chain length ($N=100$) in simulations to explore the beating frequency as a function of $\chi$. As shown in Fig10b, $\omega/\kappa_\theta \sim (\chi)^{4/3}$. This result can be understood as follows: in a full period, the energy provided by the active force, $E_f = \int_0^{2\pi/\omega} dt \sum_1^N \boldsymbol{F_a} \cdot \boldsymbol{v}$, is equal to the dissipated energy, $E_\gamma = \int_0^{2\pi/\omega} dt \sum_1^N (\boldsymbol{\gamma} \cdot \boldsymbol{v}) \cdot \boldsymbol{v}$, where $\boldsymbol{\gamma}$ is the effective friction coefficient. If using $\boldsymbol{v} \sim \omega\mathcal{R}$ and $\mathcal{R} \sim (\chi)^{-1/3}$, one can get $\omega/\kappa_\theta \sim \text{Fa}/(\gamma\mathcal{R}) \sim (\chi)^{4/3}$.

## 4. Conclusion

In summary, an experimental system composed of eccentric disks linked by a short spring, which is different from the ABP chain and tangentially-driven wormlike chain, is introduced here. The corresponding rigid-body model is proposed. Our simulations are in good agreement with experimental results, especially, qualitatively. In the free state, the short chain displays two motion modes: transverse motion and longitudinal motion. Our simulations in the zero-noise limit confirm the existence of two modes, the appearance of which is determined by the ratio of translational damping to rotational damping of eccentric disk. When tuning on the noise, the back-and-forth behavior appears at the specific rotational damping coefficient, nearly independent of stiffness. However, the back-and-forth frequency shows a non-monotonic dependence on the chain stiffness. When the chain is elongated, the chain exhibits a new stable hairpin conformation that has not been observed in SPF system.

Furthermore, when one terminal is pinned and the other is free. The whole chain rotates, almost with a constant speed,

around the pinning point. The weakly-stiff chain displays a large curvature and high speed. The rotation originates from the collective motion of disks via self-arranging their propelling direction. An interesting phenomenon that the disk closed to the pinning point simultaneously rotates around its GC was observed in experiment and simulation, which can be explained by the equation derived from the balance of force and moment of force using a simple rigid-body model. For long chain without steric interaction, it bends into a circle with a fixed radius and also rotates around the pinning point under the action of activity and rigidity. The rotational frequency scales with the flexure number as $\sim(\chi)^{4/3}$ at the small $\chi$ regime, reminiscent of the finding for the polar chain caused follower force. At the large $\chi$ regime, it scales as $\sim(\chi)^1$ because the circle's radius weakly increases with $\frac{1}{\chi}$. The different scaling exponent roots in the scaling behavior of buckling radius that accommodates the compression length. Moreover, when the terminal is clamped, the chain cannot rotates, but shows a periodic oscillation. The beating frequency also scales with the flexure number as $\sim(\chi)^{4/3}$, which can be understood via scaling analysis within the framework of balance of activity and dissipated energy.

Our experiment suggests that the macroscopic active structures can be used to mimic the dynamical behavior of living matter at the microscale where the nonlinear dynamics emerges. The work in this study is a first step towards capturing the interplay of activity, elasticity, and friction. Scientists could design other control variables to explore the new physics in the similar systems. For example, more complex geometric or topological boundaries can be used to explore the self-flapping or synchronization. In addition, the active agent here is very simple without intelligence. Further work can adopt laser-controlled robots or programmable robots through embedded technology. Then one could study the effect of translational/rotational friction or noise, independently, in experiment.

## SUPPLEMENTARY MATERIAL

See the supplementary material for additional details provided, including experimental methods and movies, rigid-body model, and methods for calculating radius of gyration, Rg, and cylindricity A.

## ACKNOWLEDGMENTS

Project supported by the National Natural Science Foundation of China of Grant Nos. 21674078 (Tian), 21774091 (Chen), and 21574096(Chen).

## AUTHOR DECLARATIONS

**Conflict of Interest**

There are no conflicts to declare.

## DATA AVAILABILITY

The data that support the findings of this study are available from the corresponding author upon reasonable request

# Constrained motion of self-propelling eccentric disks linked by a spring

Tian-liang Xu[1], Chao-ran Qin[1], Bin Tang[1], Jin-cheng Gao[1],Jiankang Zhou[2*], Kang Chen[1*], Tian Hui Zhang[1*], Wen-de Tian[1*]

[1] *Center for Soft Condensed Matter Physics and Interdisciplinary Research, Soochow University, Suzhou 215006, China*

[2]*School of Optoelectronic Science and Engineering, Soochow University, Suzhou 215006, China*

Email: tianwende@suda.edu.cn kangchen@suda.edu.cn, zhangtianhui@suda.edu.cn, health@suda.edu.cn.

## 1.Supporting Experiment

### 1.1 Single disk

The experiment setup consists of one granular robot confined by a square-shape boundary. The robot is a circular disk adhering on a Hexbug nano, which has been used to several active matter experiments[1,2]. The top view of the robot is shown in Fig.S1. The length and width of the Hexbug Nano are 4.3cm and 12.cm, respectively. Hexbug driven by a vibration motor can hop forward on a solid substrate due to its twelve flexible legs that all bends slightly backwards. The motor houses a 1.5V button cell battery that drives the vibration. We use fresh batteries in each new experiment and run experiments for less than 20 minutes to prevent battery power degrading. it has been demonstrated that the robot dynamics depends sensitively on the battery power, the robot mass, the elasticity of flexible legs, and the frictional property of the substrate [1,2].

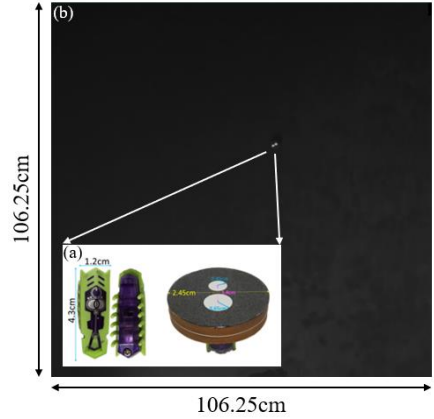

Fig. S1 (a) Hexbug and a granular robot with a disk on the back of the Hexbug. (b) The view of a robot in a square of 106.25x106.25cm.

Circular disks are made of a high-density Styrofoam sheet. The diameter of a disk is about 5cm. There are two white round spots on the disk, with diameter equal to 1.3 cm and 0.9 cm, respectively. They are used to track the position and orientation of the robot. The centers of two spots are symmetrically distributed around the center of the disk with distance of 1.4cm. The robot is driven in the direction from the large white spot to the small white spot.

We quantify robot motion via a CCD camera at a rate of 38 frames/s with a spatial resolution of $1200*1200$ pixel$^2$ over a field of view of $200 \times 200 cm^2$. Then we use MJPG software and particle-recognition code by IDL to extract the position of each white spot on the disk. We use particle-tracking algorithm based on a minimum distance criterion written by C++ code to construct trajectories from the positions. The position of each robot is calculated by $\vec{r} = \frac{\vec{r_s}+\vec{r_l}}{2}$, where $\vec{r_s}$ and $\vec{r_l}$ are the positions of small and large white spots, respectively. The orientation of each robot is given by a unit vector $\vec{n} = (\cos\theta, \sin\theta) = \frac{\vec{r_s}-\vec{r_l}}{|\vec{r_s}-\vec{r_l}|}$, where $\theta$ is the angle between the unit vector and X-axis in the in the stationary X-Y frame. From the trajectories, the instantaneous robot translational velocity is defined as $\vec{v_i} = \frac{d\vec{r_i}}{dt}$.

## 1.2 Dynamics of an isolated robot

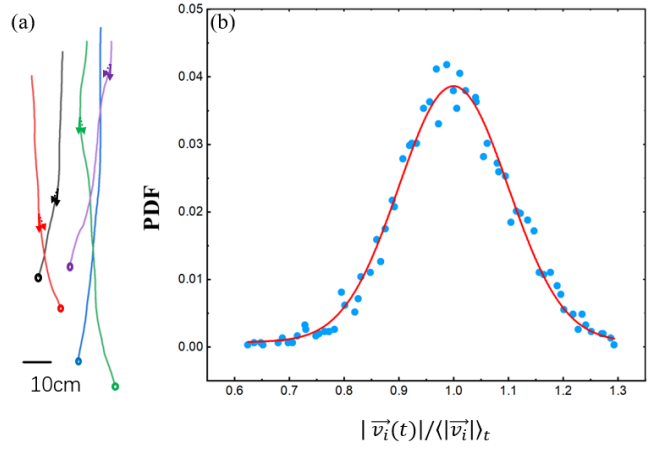

Fig.S2 (a)Trajectories of isolated robots. body orientation of robots is shown by solid arrows. (b) Probability distribution of normalized translational speed measured in six isolated robots.

Five typical trajectories of isolated robots are plotted in Fig. S2(a). These trajectories are collected in the central part of the arena to eliminate the effects of collisions between robots and the boundary. Robots mainly move in the direction of their body orientation as shown by solid arrows in panel. Trajectories consist of straight segments connected by random (but small) changes in motion direction. Robots can run persistently across the arena with small changes in their velocities and orientations. To quantify the speed fluctuations, probability distributions of $\frac{|\vec{v}(t)|}{\langle|\vec{v}(t)|\rangle_t}$ for six robots are plotted in fig. S2(b) with a Gaussian fit; the speed averaged over all times and all robots is $\bar{v} \approx 11$ cm/s. Robots change their body orientations randomly. Fluctuations in translation and orientation may arise from manufacturing imperfection of robots and surface roughness of the arena bottom. The mean square displacement of center-of-mass, MSD $(t) = \langle[\vec{r}(t+t_0) - \vec{r}(t_0)]^2\rangle$ and the mean square angle displacement of orientation, MSAD$(t) = \langle[\vec{\theta}(t+t_0) - \vec{\theta}(t_0)]^2\rangle$ of robots is given in Fig.S3(a) and (b) respectively. It shows a super-diffusion of translocation (MSD(t) $\propto t^2$) due to the self-driven ability and a normal diffusion of rotation (MSAD(t) $\propto t^1$) in the arena. The rotational diffusion is due to the vibration of legs of the motor.

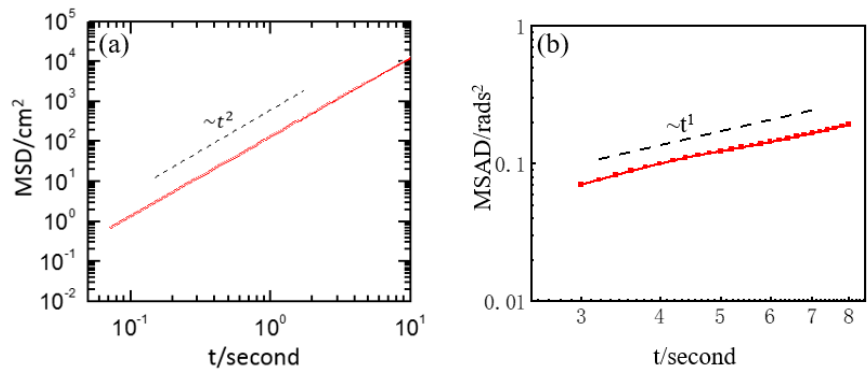

Fig.S3 The log-log plot of mean square displacement (MSD, a) and the mean square angle displacement of orientation (MSAD, b) of robots. The dash lines denote ~$t^2$ and ~$t^1$.

## 2. Supporting Figures

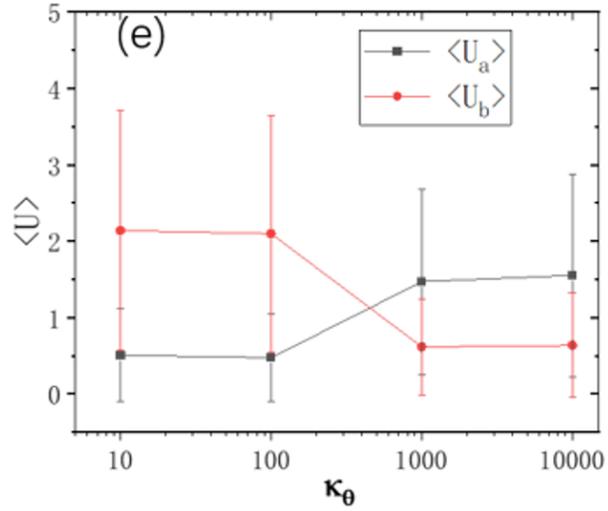

Fig.S4 Bond energy $U_b$ and bending $U_a$ energy of a chain are functions of stiffness $\kappa_\theta$ at $T_t = 1.0, \gamma_r = 1.0$. The error bar is the standard deviation data.

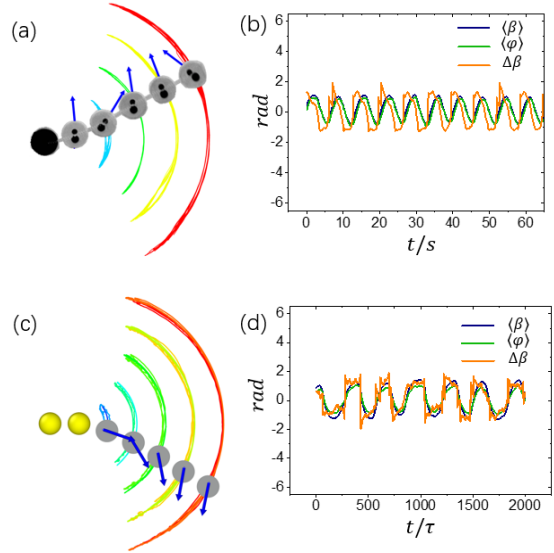

Fig.S5 (a) Snapshots of trajectories of disks for the clamped filament with strong stiffness and (c) the corresponding simulation results at $\kappa_\theta = 300$ $(left)$ with $\gamma_r = \gamma_t = 100, F = 100, T_t = 0, T_r = 1.0$. (b) The average swing angle $\langle \beta(t) \rangle$, the bending degree $\Delta\beta$, and the average disk orientation $\langle \varphi \rangle$ as a function of time obtained by experiment. (d) The corresponding simulation results shows good agreement with experimental results.

## 3. Experimental Movies

Move_S1: Lateral motion of free filament
Move_S2: Longitudinal motion of free filament
Move_S3: Rotation of pinned filament
Move_S4: Beating of clamped filament

## 4. Rigid-body model:

The schematic of our simple model is given in Fig.7b. Taking the fixed point as the coordinate origin, the motion of disk's center of mass is subject to the following geometric constraints:

$$\dot{x} = -r\dot{\psi}\sin\psi - r_l(\dot{\varphi}+\dot{\psi})\sin(\varphi+\psi) \qquad (1)$$
$$\dot{y} = r\dot{\psi}\cos\psi - r_l(\dot{\varphi}+\dot{\psi})\cos(\varphi+\psi) \qquad (2)$$

The resultant force and moment on the center of mass are:
$$m\ddot{r} = \mathbf{F} - \gamma_t \dot{r} \qquad (3)$$
$$I\ddot{\psi} = \mathbf{\Gamma} - \gamma_r \dot{\psi} \qquad (4)$$

Among them, the relationship between force and moment is as follows:
$$-F_x r_l \sin(\psi+\varphi) + F_y r_l \cos(\psi+\varphi) = \Gamma \qquad (5)$$

where $F_x$ or $F_y$ is the x/y-th component of the vector $\mathbf{F}$. We can get:
$$F_x = -\mathbf{F}\cdot\sin\psi, \quad F_x = \mathbf{F}\cdot\cos\psi \qquad (6)$$

We assume the rod rotates with a fixed angular speed $\dot{\psi} = \omega$, we get $\psi = \omega t$ and $\dot{r} = r\cdot\omega$.

Considering that the translation and rotation of the particle are overdamped, the mass and moment of inertia in equations (3) and (4) are 0 respectively. Then we get:
$$F = \gamma_t \cdot r \cdot \omega \qquad (7)$$

Substitute formula (5),(6) and (7) into formula (4), we can get:
$$\dot{\varphi} = \frac{1}{\left(\frac{\gamma_r}{\gamma_t} - r_l^2\right)}\left\{-\frac{\gamma_r}{\gamma_t}\omega + r_l^2\omega + r\omega r_l[\sin(\omega t+\varphi)\sin(\omega t) + \cos(\omega t+\varphi)\cos(\omega t)]\right\} \qquad (8)$$

Finally, we get:
$$\dot{\varphi} = \frac{1}{\left(\frac{\gamma_r}{\gamma_t} - r_l^2\right)}\left[-\frac{\gamma_r}{\gamma_t}\omega + r_l^2\omega + r\omega r_l\cos\varphi\right] = \omega\left[\frac{r}{\frac{\gamma_r}{\gamma_t r_l} - r_l}\cos\varphi - 1\right] \qquad (9)$$

Carefully analyzing the above equation, we can see that: when $r < \frac{\gamma_r}{\gamma_t r_l} - r_l$, $\dot{\varphi} < 0$ always keeps, which governs the behavior of the first disk we observed in experiment and simulation; when $r > \frac{\gamma_r}{\gamma_t r_l} - r_l$, a stationary solution, $\varphi = \mathrm{acos}\left(\frac{\frac{\gamma_r}{r_l\gamma_t} - r_l}{r}\right)$, can be obtained via setting $\dot{\varphi} = 0$, which implies that the angle is a function of $r$ and close to $\frac{\pi}{2}$ when $r \to \infty$.

## 5. Methods for calculating radius of gyration, R*g*, and cylindricity *A*

In order to further explore the influence of filament length on polymer chain conformation, 100 independent simulations were carried out for systems of *N*=20, 50, 100, 200. Radius of gyration, $Rg$ and cylindricity $A$ were calculated.
$$Rg^2 = \sum_{i=1}^{N}(\Delta x_i^2 + \Delta y_i^2) \qquad (10)$$

where, $\Delta x_i, \Delta y_i$ refers to the GC of each disk *i* relative to the GC of the filament $x,y$, respectively. To evaluate the shape, we used the radius of gyration ($Rg_0$) of a rod with the same *N* to normalize the size. It means that when the relative $\tilde{R}g = Rg/Rg_0$ is close to 1, the filament is completely straight.

To obtain the cylindricity, *A*, the rotation tensor *S* is calculated firstly:
$$S = \sum_{i=0}^{N}\begin{bmatrix}\Delta x_i\Delta x_i & \Delta x_i\Delta y_i \\ \Delta y_i\Delta y_i & \Delta y_i\Delta x_i\end{bmatrix} \qquad (11)$$
$$Tr(S) = \lambda_1 + \lambda_2$$

where $\lambda_1$ and $\lambda_2$ are two eigenvalues of the gyration tensor, respectively, and $\lambda_2 > \lambda_1$. Cylindricity is defined as:

$$A = \frac{\lambda_2 - \lambda_1}{Rg^2} \qquad (12)$$

The smaller the $A$, the closer the chain is to a circle, and the larger the $A$, the closer the chain is to a straight line.

## 6. The dimer model

The physical intuition can be uncovered by a dimer model (Fig.S6). We define $\alpha_1$ and $\alpha_2$ with range of 0~2π are angles between the active force and the end-to-end vector pointing from the 1$^{st}$ disk to 2$^{nd}$ disk, and then calculate their evolutionary tendency (d$\alpha_1$, d$\alpha_2$).

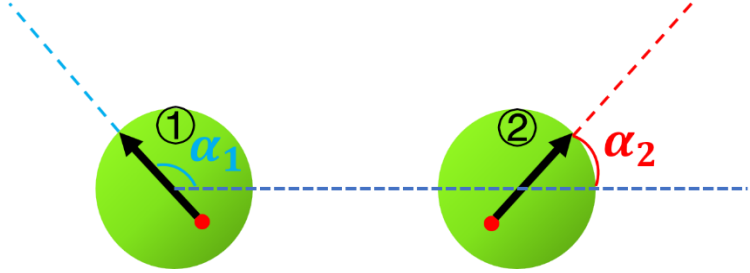

Fig.S6 The schematic of dimer model. $\alpha_1$ and $\alpha_2$ with range of 0~2π are angles between the direction active force and the end-to-end vector pointing from the 1$^{st}$ disk to 2$^{nd}$ disk.

The distance between the two disks is $r_0$. The component of the active forces parallel to the bond is $F_a(cos\alpha_1 - cos\alpha_2)$. Since the two disks are eccentric, the stress causes the disks to rotate:

$$\dot{\alpha}_1^{\parallel} = F_a \cdot sin\alpha_1 \cdot \frac{r_l}{\gamma_r}(cos\alpha_1 - cos\alpha_2)$$
$$\dot{\alpha}_2^{\parallel} = -F_a \cdot sin\alpha_2 \cdot \frac{r_l}{\gamma_r}(cos\alpha_1 - cos\alpha_2) \qquad (13)$$

The component of the active forces perpendicular to the bond can rotate the chain, causing the parameters $\alpha_1$ and $\alpha_2$ to change:

$$\dot{\alpha}_1^{\perp} = -\frac{F_a \cdot (sin\alpha_1 - sin\alpha_2)}{\gamma_t \cdot r_0}$$
$$\dot{\alpha}_2^{\perp} = -\frac{F_a \cdot (sin\alpha_1 - sin\alpha_2)}{\gamma_t \cdot r_0} \qquad (14)$$

Then, we can get the rate of time evolution of $\alpha_1$ and $\alpha_2$:

$$\dot{\alpha}_1 = \dot{\alpha}_1^{\parallel} + \dot{\alpha}_1^{\perp} = F \cdot sin\alpha_1 \cdot \frac{r_l}{\gamma_r}(cos\alpha_1 - cos\alpha_2) - \frac{F \cdot (sin\alpha_1 - sin\alpha_2)}{\gamma_t \cdot r_0}$$
$$\dot{\alpha}_2 = \dot{\alpha}_2^{\parallel} + \dot{\alpha}_2^{\perp} = -F \cdot sin\alpha_2 \cdot \frac{r_l}{\gamma_r}(cos\alpha_1 - cos\alpha_2) - \frac{F \cdot (sin\alpha_1 - sin\alpha_2)}{\gamma_t \cdot r_0} \qquad (15)$$

Similar to the main text, we set $u = \gamma_t \sigma^2 / \gamma_r$, then get

$$\dot{\alpha}_1 = \dot{\alpha}_1^{\parallel} + \dot{\alpha}_1^{\perp} = u[\frac{r_l r_0}{\sigma^2} \cdot \frac{F_a \cdot sin\alpha_1 \cdot (cos\alpha_1 - cos\alpha_2)}{\gamma_t \cdot r_0}] - \frac{F_a \cdot (sin\alpha_1 - sin\alpha_2)}{\gamma_t \cdot r_0}$$
$$\dot{\alpha}_2 = \dot{\alpha}_2^{\parallel} + \dot{\alpha}_2^{\perp} = -u[\frac{r_l r_0}{\sigma^2} \cdot \frac{F_a \cdot sin\alpha_2 \cdot (cos\alpha_1 - cos\alpha_2)}{\gamma_t \cdot r_0}] - \frac{F_a \cdot (sin\alpha_1 - sin\alpha_2)}{\gamma_t \cdot r_0} \qquad (16)$$

After choosing a short time $dt=0.001\tau$, then we can calculate the $d\alpha_1 = dt * \dot{\alpha}_1$ and $d\alpha_2 = dt * \dot{\alpha}_2$. Evidently, $d\alpha_1$ and $d\alpha_2$ are the evolutionary tendency of $\alpha_1$ and $\alpha_2$, respectively. That is to say, if we know the $\alpha_1$ and $\alpha_2$, we will know where they go. We calculate and plot the vector $(d\alpha_1, d\alpha_2)$ for all $\alpha_1 s$ and $\alpha_2 s$ in the Fig.S7 for different parameters.

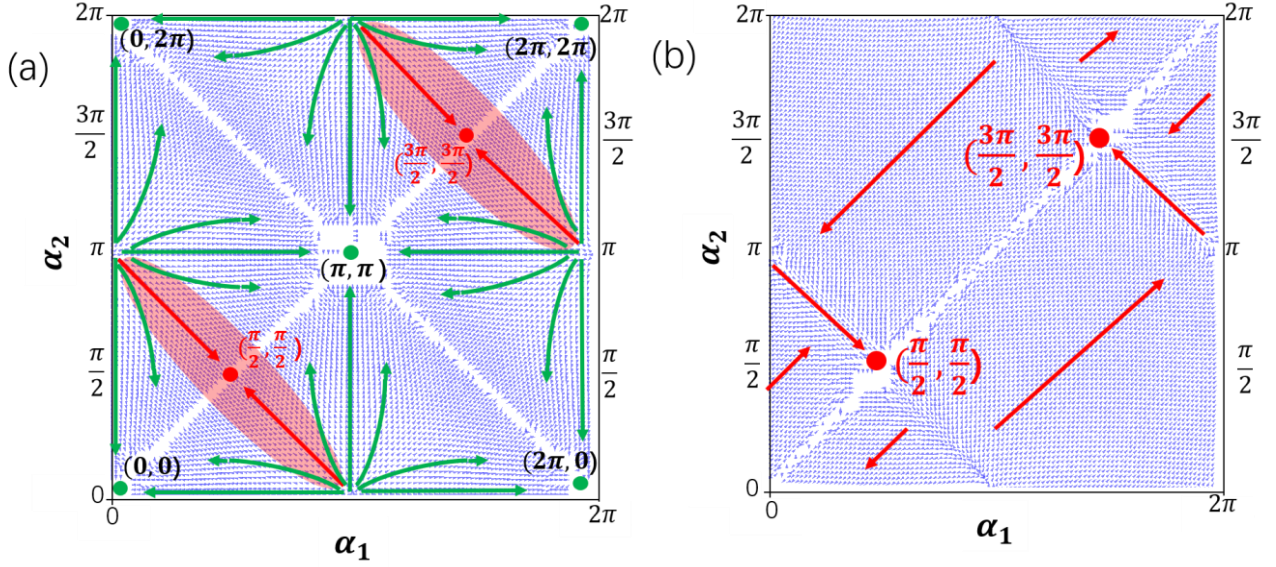

Fig. S7. The evolutionary direction of $\alpha_1$ and $\alpha_2$ for $\gamma_t = 100$, $\gamma_r = 1$ (a) and $\gamma_t = 100$, $\gamma_r = 1000$ (b). The blue arrows are vectors $(d\alpha_1, d\alpha_2)$, which denote their evolutionary tendency. The eye-guiding green/red arrows are evolutionary directions. Green and red points demonstrate the attractive centers for longitudinal and transverse mode, respectively.

FigS7.a and FigS.7b show the evolutionary direction of $\alpha_1$ and $\alpha_2$ at $u = 100$ and $u = 0.1$, respectively. It can be found that, at $u = 100$, $(\alpha_1, \alpha_2)$ prefers to $(0, 2\pi)$, $(2\pi, 0)$, $(0, 0)$, $(2\pi, 2\pi)$, and $(\pi, \pi)$, which imply the longitudinal motion. At $u = 0.1$, $(\alpha_1, \alpha_2)$ prefers to $(\frac{\pi}{2}, \frac{\pi}{2})$ and $(\frac{3\pi}{2}, \frac{3\pi}{2})$, which denotes the transverse motion.

The physical intuition in the main text can be manifested by equation 16. When $u \gg 1$, the first term ($\dot{\alpha}^{\parallel}$) is dominant; while $u \ll 1$ the second term ($\dot{\alpha}^{\perp}$) is dominant. Acoording to our analysis, the first term triggers the rotation of disks and the second term induces the rotation of chain.

## 7. The method to estimate the rigidity of spring

We use the method as shown in the schematic diagram below to estimate the spring rigidity. One end of the spring is clamped, and the force F on the other end causes the spring to bend to the dotted line. The bending stiffness was calculated using $\frac{FL^2}{2\theta}$.

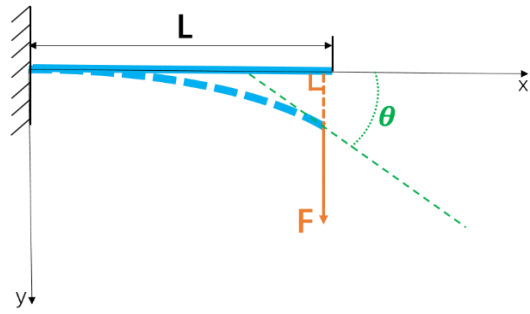

Fig.S8: The schematic for measuring stiffness of spring